\documentclass[10pt,prl,aps,twocolumn,superscriptaddress,floatfix,showpacs]{revtex4}
\usepackage{graphicx}
\usepackage{amssymb}
\usepackage{bm}

\begin{document}

\title{Variational quantum Monte Carlo simulations with tensor-network states}

\author{A. W. Sandvik} 
\affiliation{Department of Physics, Boston University, 590 Commonwealth Avenue, Boston, Massachusetts 02215}
\affiliation{Department of Physics, National Taiwan University, Taipei, Taiwan 106}

\author{G. Vidal} 
\affiliation{School of Physical Sciences, The University of Queensland, QLD 4072, Australia}

\date{\today}

\begin{abstract}
We show that the formalism of tensor-network states, such as the matrix product states (MPS), can be used as a 
basis for variational quantum Monte Carlo simulations. Using a stochastic optimization method, we demonstrate 
the potential of this approach by explicit MPS calculations for the transverse Ising chain with up to $N=256$ spins 
at criticality, using periodic boundary conditions and $D\times D$ matrices with $D$ up to $48$. The computational 
cost of our scheme formally scales as $ND^3$, whereas standard MPS approaches and the related density matrix 
renormalization group method scale as $ND^5$ and $ND^6$, respectively, for periodic systems. 
\end{abstract}


\maketitle

Devising unbiased computational methods for correlated quantum many-body systems remains one of the greatest 
challenges in theoretical physics. Considerable progress has been made in recent years. Quantum Monte Carlo 
(QMC) methods with efficient loop-cluster updates \cite{evertz,prokofev,directed} now enable simulations of 
certain classes of spin and boson hamiltonians on very large lattices---up to $>10^4$ sites essentially in the 
ground state and considerably more at elevated temperatures. Modern projector QMC methods \cite{vbmethod} 
can also access large lattices. Both approaches are already contributing
significantly to forefront areas of condensed matter physics, e.g., studies of exotic quantum phase transitions
in antiferromagnets \cite{deconf}. However, due to "sign problems" \cite{loh,henelius}, most fermion 
systems in more than one dimension and spin models with frustrated interactions are intractable to QMC simulations. 
The density matrix renormalization group (DMRG) method \cite{white,schollwock}, on the other hand, can produce 
essentially exact results for one-dimensional fermion systems and frustrated spins, including systems of a few 
coupled chains (ladders) \cite{ladders}. These calculations are often restricted to open boundary conditions,
however, which sometimes can be problematic. A more severe limitation is the exponential scaling 
in the computational complexity for systems with two or more dimensions \cite{liang}.

The underlying reason for the problems with DMRG in higher dimensions has recently been identified as the inability 
of matrix product states (MPS), which are produced by the DMRG method \cite{ostlund}, to properly account for 
entanglement in dimensions higher than one \cite{vidal1}. In order to overcome this limitation, a generalization 
of the MPS was proposed---the projected-entangled pair states (PEPS) \cite{peps}. These states are based on 
tensor-product networks \cite{nishino}, which are contracted using an approximate scheme. While this approach is very 
promising, practical applications are still hampered by the severe increase of the computational effort with the size $D$ of 
the tensors in two dimensions. The scaling is typically $\sim{D^{12}}$, and calculations are therefore currently 
restricted to very small $D \sim 2-5$ \cite{isacsson,murg,jordan}. Developing schemes with a more favorable $D$ scaling 
is therefore a high priority.

In principle MPS and PEPS can be used in variational QMC calculations. Sampling the physical states, instead of 
contracting the tensor network over those indices, formally reduces the scaling in $D$ \cite{scaling}. In practice, 
it is not clear how much can be achieved this way, however. An efficient method is required to optimize tensors 
with hundreds or thousands of independent parameters, based on noisy Monte Carlo estimates of the energy and its 
derivatives. In this Letter we demonstrate that such a program is actually feasible. We develop a method based on 
a stochastic optimization scheme \cite{jievar} which requires only the first energy derivatives. Here we focus on 
MPS for simplicity, but the scheme can be applied to more generic tensor networks, e.g., PEPS, as well. We test 
the method on the Ising chain in a transverse external field,
\begin{equation}
H = -\sum_{i=1}^L (\sigma^z_i \sigma^z_{i+1} + h\sigma^x_i),
\end{equation}
where $\sigma^x$ and $\sigma^z$ are the standard Pauli matrices. This system undergoes a quantum phase 
transition from a ground state with long-range Ising order in the $z$ direction for $h < 1$ to a state with 
disordered $z$ components when $h > 1$. We here consider exclusively the computationally most challenging 
$h=1$ critical point.

For a periodic chain, a translationally invariant matrix-product state 
with momentum $k=0$ is of the form \cite{ostlund}
\begin{equation}
|\Psi\rangle = \sum_{\{s\}} {\rm Tr}\{ A(s_1)A(s_2)\cdots A(s_N)\}
|s_1,s_2,\ldots,s_{N}\rangle ,
\label{mpstate}
\end{equation}
where the spins $s_i = \pm 1$ are the eigenvalues of $\sigma^z_i$ and $A(\pm 1)$ are two $D\times D$ matrices 
(for a non-translationally invariant system the matrices would be site dependent). We here take the matrices to be 
real and symmetric, which, from properties of the trace, corresponds to a $s_i \to s_{N-i+1}$ reflection
symmetric state. The ground state should also be invariant with respect to spin inversion; $s_i \to -s_i$ for 
all $i$. A sufficient condition for this is that $A(\pm 1)$ are related by a transformation $U$ such 
that $U^{-1}A(1)U=A(-1)$ and $U^{-1}A(-1)U=A(1)$, which implies $U^2=I$ (the identity matrix). For simplicity, 
and because of indications that a greater flexibility of the matrices is advantageous for the optimization, 
we here only enforce the weaker condition that $A(1)$ and $A(-1)$ have identical eigenvalues, using a scheme 
discussed below.

Our goal is to find the matrix elements $a_{ij}^s$, $s=\pm 1$, that minimize the MPS energy
$E=\langle H\rangle$. Denoting the wave function coefficient for state $|S\rangle=|s_1,\ldots,s_N\rangle$
\begin{equation}
W(S) = {\rm Tr}\{ A(s_1)A(s_2)\cdots A(s_N)\},
\label{ws}
\end{equation}
the energy, for given matrices $A(\pm 1)$, can be written in the form appropriate
for Monte Carlo sampling;
\begin{equation}
E = \frac{1}{Z}\sum_S W^2(S)E(S),~~~ Z=\sum_S W^2(S), 
\end{equation}
where $E(S)$ is the estimator
\begin{equation}
E(S) = \sum_{S'} \frac{W(S')}{W(S)}\langle S'|H|S\rangle.
\label{es}
\end{equation}
The energy  
can be evaluated using importance sampling of the spin configurations according to the weight $W^2(S)$; 
$E = \langle E(S)\rangle$. Our scheme also requires the derivatives of the energy with respect to 
the matrix elements;
\begin{equation}
\frac{\partial E}{\partial a^s_{ij}}=
2\langle \Delta^s_{ij}(S)E(S)\rangle - 2\langle \Delta^s_{ij}(S)\rangle\langle E(S)\rangle,
\label{ederiv}
\end{equation}
where we have defined
\begin{equation}
\Delta_{ij}^s= \frac{1}{W(S)}\frac{\partial W(S)}{\partial a_{ij}^s}.
\end{equation}
Introducing the matrices
\begin{equation}
B(m)=A(s_{m+1})\cdots A(s_N) A(s_1)\cdots A(s_{m-1}),
\label{bm}
\end{equation}
the derivative of the weight (\ref{ws}) is
\begin{equation}
\frac{\partial W(S)}{\partial a_{ij}^s}= 
\frac{1}{1+\delta_{ij}}\sum_{m=1}^N [B_{ij}(m) + B_{ji}(m)]\delta_{s,s_m}.
\label{derij}
\end{equation}

We sample the states by generating successive configurations from a stored $S$ by single-spin flips; 
$s_m \to -s_m$. We denote the new tentative configuration $S'_m$. Visiting the spins sequentially; 
$m=1,2,\ldots,N$, we flip them according to the Metropolis probability; 
$P_{\rm flip}={\rm min}[W^2(S'_m)/W^2(S),1]$. To evaluate $P_{\rm flip}$, we use the cyclic 
property of the trace and write the new coefficient as $W(S'_m) = {\rm Tr}\{ A(-s_m)B(m)\}$. Further, we write the 
matrix $B(m)$ in Eq.~(\ref{bm}) as a product of left and right matrices $B(m)=L(m+1)R(m-1)$, where 
$L(m)=A(s_{m})\cdots A(s_N)$ and $R(m)=A(s_1)\cdots A(s_{m})$. We also define $L(N+1)=R(0)=I$. Before 
starting the updating process we calculate and store the left matrices $L(2),\ldots,L(N)$, based on the
initial spin configuration (random or from a previous run). Each successive spin-flip attempt then requires 
only one matrix multiplication, and another for advancing the right matrix; $R(m)=R(m-1)A(s_m)$. Since 
$L(m)$ is no longer needed at this stage we store $R(m)$ in its place for future use.

Diagonal quantities, e.g., the Ising part of the energy, $E_z=-\sum_i \sigma^z_i\sigma^z_{i+1}$, 
can be simply obtained by averaging the appropriate spin 
correlations in the stored state $|S\rangle$. To calculate off-diagonal quantities, ratios $W(S')/W(S)$ 
are needed. After a full sweep of spin updates, all the matrices $R(m)$ have been generated and stored. We can use 
them to speedily measure the off-diagonal energy $E_x=h\sum_i\langle \sigma^+_i  + \sigma^-_i \rangle$, 
the estimator of which is
\begin{equation}
E_x(S) = h\sum_{m=1}^N \frac{W(S'_m)}{W(S)},
\end{equation}
as well as the derivatives (\ref{derij}).  To evaluate the sums, we now traverse the system from $m=N$ to $1$, 
and in the process generate the left matrices $L(m)$ and store them in the place of $R(m)$.  
Once this process is completed we again have what we need to carry out an updating sweep 
in the manner described above. A full updating sweep, including measurements, thus requires $4 N$ matrix 
multiplications (plus operations which have a lower scaling in $D$), giving a formal scaling $ND^3$ of
the algorithm.

Carrying out successive simulations with fixed matrices $A(\pm 1)$, the energy and derivatives obtained 
on the basis of some number $F$ of spin-flip sweeps (referred to as one simulation {\it bin}) are used to 
update the matrix elements with $j\ge i$ according to (and subsequently $a_{ji}^s=a_{ij}^s$)
\cite{jievar}:
\begin{equation}
a_{ij}^s \to a_{ij}^s - \delta(k)\cdot r_{ij}^s\cdot {\rm sign}(\partial E/\partial a^s_{ij}).
\end{equation}
Here $r_{ij}^s \in [0,1)$ is random and $\delta (k)$ is the maximum change, which decreases as a function of 
a counter $k=0,1,..$. Thus, instead of moving in the direction of the approximately evaluated gradient, as in 
standard stochastic optimization \cite{stocopt,harju}, each parameter is changed independently, using the 
``correct'' sign but with a random and well bounded magnitude for the step. This results in a very stable 
optimization ideally suited for problems with large numbers of parameters.
For the gradual reduction of $\delta$, we here use a geometric form; $\delta = \delta_0 Q^k$, with,
typically, $Q=0.9-0.95$, but other forms also work well, e.g., $\delta =\delta_0 k^{-\alpha}$, with 
$\alpha \in [1/2,1]$. For each $k$, we complete a number, $G$, of bins, each followed by updates of the 
matrix elements. The number of sweeps per bin, $F$, as well as $G$ are increased with $k$. The rationale behind 
increasing $F(k)$ is that, as we approach the energy minimum, the derivatives will become smaller and require 
more sampling in order not to be dominated by noise \cite{noisenote}. Increasing $G$ leads effectively 
to a  slower ``cooling'' rate. We typically use a linear dependence in both cases; $F=F_0k$, $G=G_0k$. We 
output the energy and its statistical error computed on the basis of the $G$ bins before each increment of $k$. 
Since $F$ and $G$ increase with $k$,  the error bars will decrease. For a sufficiently long run, if the cooling 
is slow enough, the calculated $E_D(k)$ should approach the optimal energy for a given matrix size $D$.

\begin{figure}
\includegraphics[width=7cm, clip]{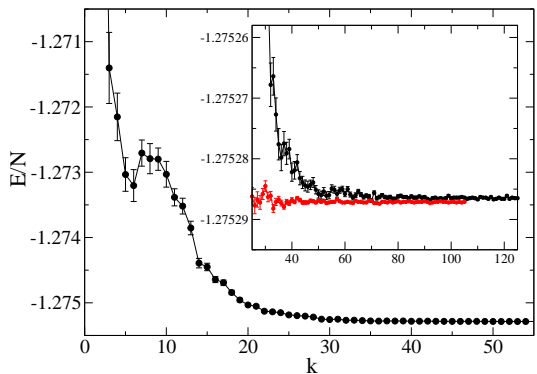}
\caption{(Color online) Main panel: Convergence of the energy per site of a $16$-site system at $h=1$, 
using $D=8$ and starting from random  matrices. The cooling parameters were $Q=0.9$, $\delta_0=0.05$, $G_0=10$, 
$F_0=100$. Inset: The later stages of the simulation on a more detailed scale, and a comparison with a run which 
started from a converged $D=6$ calculation (lower curve; red symbols); here $\delta_0=0.005$, $G_0=5$, $F_0=50$.}
\label{fig1}
\vskip-3mm
\end{figure}

As we already mentioned, we wish to enforce the property that $A(-1)$ and $A(1)$ have the same eigenvalues.
We do this after each adjustment of the matrix elements, by diagonalizing both matrices and averaging their
eigenvalues. The averaged diagonal matrix is then transformed back using the diagonalizing matrices for the 
original $A(\pm 1)$. If we do not carry out this diagonalization step we still in practice do obtain matrices 
with approximately equal eigenvalue spectra. However, enforcing this condition exactly seems to have favorable 
effects on the ability of the optimization method to quickly converge to a spin-inversion invariant ground 
state. We normalize the matrices so that the largest element $|a_{ij}^s|=1$.

It should be noted that the optimal matrices are not unique---there is a huge degeneracy in terms of 
simultaneous transformations of $A(\pm 1)$ that leave the trace invariant. This may also be an advantage in the
optimization, as we are not trying to locate a point, but only reach some large hypersurface 
in parameter space.

In Fig.~\ref{fig1} we show an example of the convergence of the optimization for a $16$-site chain,
using $D=8$ and starting from random $A(\pm 1)$. The initial maximum parameter shift was $\delta_0=0.05$. 
We compare with a run which started from matrices resulting from a calculation with $D=6$ (with the new matrix 
elements in the larger matrices generated at random in the range $[-\delta_0,\delta_0$]), which allows for a 
smaller initial step $\delta_0=0.005$. The latter calculation produces a marginally lower energy, showing that 
the cooling rate in the former case was slightly too fast---cooling slower we obtain consistent results. 

It is useful to start the optimization for some $N$ and $D$ from $A(\pm 1)$ previously obtained for a smaller $N$ 
and the same $D$, or the same $N$ and smaller $D$. Another good strategy is to first do a short run with a large 
$\delta_0 \approx 0.1$ to achieve convergence only approximately, and then to restart the calculation with a 
smaller $\delta_0$ [but much larger than the smallest $\delta(k)$ reached previously]. After a few such restarts
there are typically no further changes in the minimum energy reached.

We do not claim that the cooling protocol presented above is optimal; further improvements could potentially lead to 
considerable efficiency gains. However, even as it stands now the scheme performs remarkably well.

\begin{figure}
\includegraphics[width=6.25cm, clip]{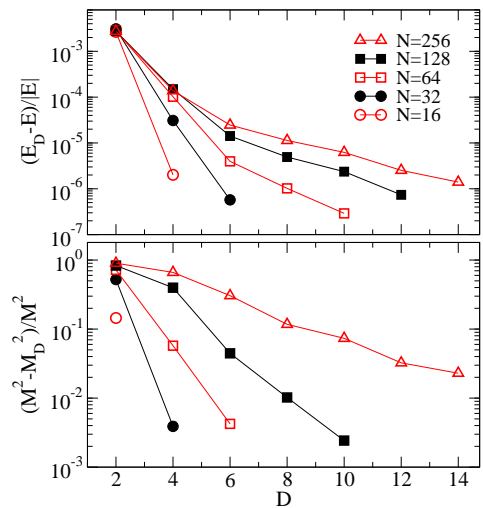}
\caption{(Color online) Relative error of the energy and squared magnetization versus
the matrix dimension $D$.}
\label{fig2}
\vskip-3mm
\end{figure}

We now compare simulation results with the exact solution \cite{exact} of the critical transverse Ising chain. 
we consider the energy as well as the squared magnetization; $M^2=(\sum_i \sigma^z_i)^2/N^2$. The convergence 
with $D$ is illustrated in Fig.~\ref{fig2}. In the case of the energy, a desired relative accuracy requires a $D$ which
eventually approaches a constant for large $N$. The squared magnetization is directly related to the long-distance
physics, however, and our results are consistent with the expectation that $D$ has to grow as some power, 
$D \sim N^\alpha,$ to achieve a given relative accuracy. From Fig.~\ref{fig2} we obtain, roughly, $\alpha$ in the
range $0.5 - 1$. The statistical errors in Fig.~\ref{fig2} are smaller than the symbols. The slight jaggedness of the 
curves for $L=256$, in particular, reflects the fact that it is not possible in practice to reach the optimum exactly. 
Nevertheless, it is clear from these tests that our scheme allows for a systematic approach to the ground state.  

\begin{table}
\caption{Variational QMC results for the critical transverse Ising chain compared with the exact solution \cite{exact}. 
The error bars of the MPS energies are $\approx 10^{-8}$ or smaller.}
\label{tab1}
\begin{ruledtabular}
\begin{tabular}{llllll}
$N$     & $D$ & $E/N$~(MPS)   & $E/N$~(ex.) &  $M^2$~(MPS)  &  $M^2$~(ex.)   \\  
\hline
16     & 12 &  $-1.27528715$    &  $-1.27528715$  &  $0.52233(2)$  &  $0.522332$   \\  
32     & 16 &  $-1.27375097$    &  $-1.27375102$  &  $0.44076(5)$  &  $0.440795$   \\  
64     & 20 &  $-1.27336736$    &  $-1.27336739$  &  $0.37151(9)$  &  $0.371455$   \\  
128    & 32 &  $-1.27327145$    &  $-1.27327150$  &  $0.3126(1)$   &  $0.312752$   \\  
256    & 48 &  $-1.27324731$    &  $-1.27324753$  &  $0.2630(2)$   &  $0.263192$   \\  
\end{tabular}
\end{ruledtabular}
\vskip-3mm
\end{table}

In Table \ref{tab1} we show results for the largest $D$ considered for each $N$. The statistical errors of the energies 
are not shown, but are at most $\pm 2$ in the last digit (i.e., $10^{-8}$). For a variational wave function that can exactly 
reproduce the exact ground state, which should be the case here for $D \to \infty$, the fluctuations in the energy should 
vanish. We indeed observe a strong reduction of the statistical errors of $E_D(k)$ with increasing $D$, as reflected in the
very small error bars. For $N > 16$, there is still some small discrepancies beyond statistical errors, which we believe
are not due to the finite $D$ but incomplete optimization. The ability of a stochastic scheme to reach 
so close to the optimum is still quite remarkable.

We have also carried out simulations with general non-symmetric matrices. In order to strictly enforce
the lattice reflection and spin-inversion symmetries, we then use a wave function with a trace of four different 
matrix products related to each other by these symmetries, i.e., 
\begin{equation}
W(S) = {\rm Tr}\{P(S)+P(S_R)+P(-S)+P(-S_R)\},
\label{mpstate4}
\end{equation}
where $P(S)=A(s_1)\cdots A(s_N)$, and $-S$ and $S_R$ are obtained by, respectively, spin-inverting 
and reflecting the configuration $S$. For given $D$, this wave function has a lower optimum energy
than one with a single product $P(S)$. The energy is also better than for the symmetric matrices 
discussed above. The computational effort is higher by a factor of $4$, however, and the optimization 
converges slightly slower. 

In summary, we have demonstrated that the variational QMC approach can be successfully combined with the versatility 
of tensor-network states, for a sign-problem free and systematically refinable (through the tensor dimension $D$)
generic many-body method. The scaling with the matrix size $D$ in the case of MPS for periodic chains is formally 
reduced from $D^5$ \cite{porras} to $D^3$, and similar reductions are possible with tensor networks in higher dimensions
\cite{scaling}. There may of course be some further non-obvious $D$ dependence in the convergence properties of the 
sampling and optimization schemes---its is clear that stochastic optimization will be difficult in practice for $D$
much larger than the maximum $D=48$ considered here. It should be noted, however, that other MPS schemes, as well as 
DMRG, also have convergence issues beyond the formal scaling in $N$ and $D$.

At the late stages of completing this work we became aware of Ref.~\cite{string}, where a different QMC
approach is proposed in the same spirit and applied to ``string'' states.

AWS would like to thank Y.-J. Kao for stimulating discussions. This work was supported by the NSF under 
grant No.~DMR-0513930 (AWS) and by the Australian Research Council grant No.~FF0668731 (GV). AWS also
gratefully acknowledges support from the National Center for Theoretical Sciences, Hsinchu, Taiwan.

\null


\vskip-9mm
\begin{thebibliography}{00}

\bibitem{evertz}
H. G. Evertz, Adv. Phys. {\bf 52}, 1 (2003).

\bibitem{prokofev}
N. V. Prokof\'ev, B. V. Svistunov, and I. S. Tupitsyn, Zh. Eks. Teor. Fiz. {\bf 114}, 570 (1998) 
[JETP {\bf 87}, 311 (1998)].

\bibitem{directed}
O. F. Sylju{\aa}sen and A. W. Sandvik, Phys. Rev. E {\bf 66}, 046701 (2002). 

\bibitem{vbmethod}
A. W. Sandvik, Phys. Rev. Lett. {\bf 95}, 207203 (2005).

\bibitem{deconf}
A. W. Sandvik, Phys. Rev. Lett. {\bf 98}, 227202 (2007); R. G. Melko and R. K. Kaul, 
ArXiv:0707.2961; K. Harada, N. Kawashima, and M. Troyer, ArXiv:cond-mat/0608446.

\bibitem{loh}
E. Y. Loh {\it et al.}, Phys. Rev. B {\bf 41}, 9301 (1990).

\bibitem{henelius}
P. Henelius and A. W. Sandvik, Phys. Rev. B {\bf 62}, 1102 (2000).

\bibitem{white}
S. R. White, Phys. Rev. Lett. {\bf 69}, 2863 (1992).

\bibitem{schollwock}
U. Schollw\"ock, Rev. Mod. Phys. {\bf 77}, 259 (2005).

\bibitem{ladders}
S. R. White and D. J. Scalapino, Phys. Rev. Lett. {\bf 91}, 136403 (2003).

\bibitem{liang}
S. Liang and H. Pang, Phys. Rev. B {\bf 49}, 9214 (1994).

\bibitem{ostlund}
S. \"Ostlund and S. Rommer, Phys. Rev. Lett. {\bf 75}, 3537 (1995).

\bibitem{vidal1}
G. Vidal, J. I. Latorre, E. Rico, and A. Kitaev, Phys. Rev. Lett. {\bf 90}, 227902 (2003). 

\bibitem{peps}
F. Verstraete and J. I. Cirac, Arxiv:cond-mat/0407066.

\bibitem{nishino}
T. Nishino {\it et al.}, Nucl. Phys. B {\bf 575}, 504 (2000).

\bibitem{isacsson}
A. Isacsson and O. F. Sylju{\aa}sen, Phys. Rev. E {\bf 74}, 026701 (2006).

\bibitem{murg}
V. Murg, F. Verstraete, and J. I. Cirac, Phys. Rev. A {\bf 75}, 033605 (2007).

\bibitem{jordan}
J. Jordan, R. Or\'us, G. Vidal, F. Verstraete, and J. I. Cirac,
ArXiv:cond-mat/0703788.

\bibitem{scaling}
The scaling with MPS is reduced from $D^3$ and $D^5$ \cite{porras} for open and periodic boundaries, 
respectively, to  $D^2$ and $D^3$. For PEPS with open boundaries the sacling goes from 
$D^8 \tilde{D}^2 \approx D^{12}$ \cite{murg} to  $D^{4} D'^{2}+D^2 D'^{3} \approx D^6$, where 
$\tilde{D} \approx D^2$ and $D'\approx D$ are the ranks of the boundary MPS employed in the 
contraction. A reduction of several powers of $D$ is also achieved in the case of PEPS 
with cylinder and torus boundary conditions.  

\bibitem{jievar}
J. Lou and A. W. Sandvik, Phys. Rev. B {\bf 76}, 104432 (2007).

\bibitem{stocopt}
H. Robbins and S. Monro, Ann. Math. Stat. {\bf 22}, 400 (1951);
J. C. Spall, in {\it Wiley Encyclopedia of Electrical and Electronics Engineering,
Vol.~20}, Edited by J. G. Webster (Wiley, 1999).

\bibitem{harju}
A. Harju, B. Barbiellini, S. Siljam\"aki, R. M. Nieminen, and G. Ortiz,
Phys. Rev. Lett. \textbf{79}, 1173 (1997).

\bibitem{noisenote}
Stochastic optimization takes advantage of noise \cite{stocopt}, but there is some 
limit beyond which too many errors in the signs of the derivatives are detrimental.

\bibitem{exact}
T.W. Burkhardt and I. Guim, J. Phys. A {\bf 18}, L33 (1985);
T. D. Shultz, D. C. Mattis and E. H. Lieb, Rev. Mod. Phys. {\bf 36}, 856 (1964).

\bibitem{porras}
F. Verstraete, D. Porras, J. I. Cirac  Phys. Rev. Lett. {\bf 93}, 227205 (2004).

\bibitem{string}
N. Schuch, M. M. Wolf, F. Verstraete, and J. I. Cirac, ArXiv:0708.1567.

\end{thebibliography}
\end{document}